\documentclass[preprint2]{aastex}
\usepackage{spr-astr-addons}
\usepackage{natbib}
\usepackage{graphicx}
\usepackage{latexsym}

\newcommand{\yr}{\,{\rm yr}}
\newcommand{\e}{{\rm e}}
\newcommand{\kms}{\,{\rm km\,s}^{-1}}
\begin{document}
\title{ AGN effect on cooling flow dynamics }

\shorttitle{The impact of AGN on cooling flows}
\shortauthors{F. Alouani Bibi et al.}

\author{F. Alouani Bibi, J. Binney, K. Blundell and H. Omma}
\affil{St Johns Research Centre \, University of Oxford, Oxford OX1 3JP, UK}
\email{alouani@astro.ox.ac.uk}

\begin{abstract}
 We analyzed the feedback of AGN jets on cooling flow clusters using
three-dimensional AMR hydrodynamic simulations.  We studied the interaction
of the jet with the intracluster medium and creation of low X-ray emission
cavities (Bubbles) in cluster plasma. The distribution of energy input by
the jet into the system was quantified in its different forms, i.e.
internal, kinetic and potential. We find that the energy associated with the
bubbles, $(pV+\gamma pV/(\gamma-1))$, accounts for less than 10 percent of the
jet energy.
\end{abstract}


\section{Introduction}
\label{intro}

During the last
two decades clusters with cooling flows have been extensively  studied. Prior to results obtained by XMM-Newton and Chandra it was
widely believed that in these systems up to $1000M_\odot\yr^{-1}$ of gas was
moving from the hot phase $T>10^7\,$K to low temperatures. In this picture
extensive emission by Fe XVII ions was predicted, in contradiction with
spectra obtained by XMM-Newton  \citep{Peterson03}. The reasons for this
conflict between theory and observation are discussed  in papers by
Omma et .al (2004), Binney (2005) and Pizzalato \& Soker (2005).

It is now widely accepted that feedback from AGN jets is responsible for
slowing or halting the cooling of gas in galaxy groups and clusters
\citep{Churazov,Birzan,Binneyetal07}, and it is likely  that this
phenomenon also takes place in elliptical galaxies at smaller (spatial and
temporal) scales \citep{BinneyTabor}. 

AGN jets reheat hot gas by doing mechanical work during the inflation of
low-density bubbles.  The Chandra observatory has now detected bubbles as
regions of low X-ray emission in a large number of systems
\citep{Carilli,David,McNamara,Fabian}.  Most, but not all  bubbles
are associated with non-thermal radio emission \citep{Birzan}, and when both
radio-loud and radio-quiet bubbles occur, the radio-quiet bubbles are
further from the center. Thus it appears that bubbles are radio-loud for a
period after inflation, and a combination of adiabatic expansion of the
plasma and radiative losses by the most energetic electrons makes the bubble
radio-quiet. 

In some cases a synchrotron-emitting jet runs from the galactic nucleus to
the bubbles, so jets are definitely involved in bubble formation. It is
unclear to what extent bubbles are driven by the relativistic cores of the
jets rather than the slower and more massive flows that probably surround the
cores.  We are also uncertain whether the jets are driven by the black
hole's accretion disk or emerge from its ergosphere \citep{Blandford}, but
the observation of fast jets in microquasars that are known to be powered by
neutron stars suggests that an ergosphere does not need to be involved
\citep{NipotiBB}.

\section{Method}
\label{sec:1}

To investigate how an AGN jet interacts with cluster gas, we performed a
series of three-dimensional hydrodynamic simulations using the code Enzo
\citep{Bryan}. Because of the inhomogeneity of the plasma, we used the
adaptive grid refinement (AMR) technique to subdivide the simulation domain,
which was a cube $L = 614 \,$kpc on a side. Our coarsest grid has 16 cells
in each direction, and in the $l^{\rm th}$ level of refinement the effective number
of grid points is increased to $16 \times 2^l$, where $l=0,\ldots,6$.  The
cluster core is at the 6$^{\rm th}$ refinement level, so the grid would have
$1024^3$ cells if the entire domain were refined to this level.  We used the
piecewise parabolic method PPM \citep{Woodward} Riemann solver to advance
the hydrodynamic equations.

In previous simulations the effects of jets have been taken into account by
one of three techniques: (i) an inner boundary is introduced on which inflow
boundary conditions can be set \citep{BassonA}; (ii) spherical cavities are
carved out of the ambient medium and filled with much less dense material
that provides a similar pressure \citep{Churazov01}; (iii) at some
arbitrarily chosen locations, energy is added at some specified rate
\citep{BruggenK}. Our technique differs from approach (iii) in that we add
mass, momentum and energy on two small, centrally located ``injection
disks'' of radius $r_{\rm jet}=3\,$kpc \citep{Omma}.  The ratio of the rates
of injection of mass, $\dot m_{\rm jet}=1M_\odot\yr^{-1}$, and momentum
specify the jet speed $v_{\rm jet} = 4 \times 10^4\kms$, which is
sub-relativistic but supersonic.  As the simulation develops, the region in
which the injected energy is largely thermalized moves away from the center,
just as the hot spot of a real system should. Hence we believe our technique
is more realistic than either technique (ii) or technique (iii), and it
avoids the introduction of the unphysical inner boundary required by
technique (i).  We implement this scheme by solving the equations
 \begin{equation}
\frac{\partial \rho} {\partial t} 
+ \overrightarrow{\nabla} \cdot \rho \overrightarrow{v} = 
\chi \dot m_{\rm jet} \\
\end{equation}
\begin{equation}
\frac{\partial \rho \overrightarrow{v}} {\partial t} + 
\overrightarrow{\nabla} \cdot(\rho \overrightarrow{v} 
\otimes \overrightarrow{v})
=  -\overrightarrow{\nabla} p  - \rho \overrightarrow{\nabla} \Phi
+ \chi \dot m_{\rm jet} v_{\rm jet} \overrightarrow{i}\\
\end{equation}
\begin{equation}
\frac{\partial E} {\partial t} + \overrightarrow{\nabla} \cdot (E 
\overrightarrow{v}) 
= -\overrightarrow{\nabla} \cdot p \overrightarrow{v} 
- \rho \overrightarrow{v} \cdot \overrightarrow{\nabla} \Phi
+ \chi E_{\rm jet} - j_{\rm rad}
\end{equation}
 where $\rho$, $v$, $p$, and $E$ are the mass density, velocity, thermal
pressure and total energy, respectively, $\Phi$ is the gravitational
potential, $E_{\rm jet}=\frac12\dot mv_{\rm jet}^2$, $j_{\rm rad}$ is rate
per unit volume of radiative energy loss, and $\chi$ is a window function
that vanishes outside the injection disk. On the disk 
 \begin{equation}
\chi (r) = \begin{cases}
 \e^{ - 3r^2 /2r_{\rm jet}^2 }&r \le r_{\rm jet}\\
 0&r > r_{\rm jet},
\end{cases}
\end{equation}
 where $r$ is distance from the jet axis. The cluster baryonic matter is
assumed to evolve in a dark-matter halo, described by the NFW model
\citep{Navarro}. The mass within a sphere of radius $r$ is
 \begin{equation}
M(r) = 4\pi \delta _c \rho _c r_s^3\left[ {\ln (1 + r/r_s ) - \frac{{r/r_s }}{{1 + r/r_s }}}\right],
\end{equation}
 where $\rho_c$ is the critical density of the universe at the cluster
redshift, $r_s$ is the scale radius of the cluster and $\delta_c$
is the overdensity parameter \citep{David}
\begin{eqnarray}
 \rho _c & =& \frac{{3H^2 }}{{8\pi G}} \\ 
\delta _c & =& \frac{{200}}{3}\frac{{(r_{200} /r_s )^3 }}{{\ln (1 + r_{200} /r_s ) - \frac{{r_{200} /r_s }}{{1 + r_{200} /r_s }}}},
\end{eqnarray} 
 where $r_{200}$ is the radius within which the average density is equal to
$200 \rho_c$. In our simulations $\delta_c$ and $r_s$ were equal to
$7.48 \times 10^4$ and $77\,$kpc, respectively.
 The initial electron density was
\begin{equation}
n_e (r) = 0.1\left(1+ \frac{r}{50\,{\rm kpc}}\right)^{-2.7}{\rm cm}^{-3}.
\end{equation}
 This provides a good fit to the Hydra cluster density profile reported by
\cite{David}. As initial condition we assume that the system is in
hydrostatic equilibrium and deduce the initial temperature profile from the
assumption that the plasma is an ideal gas.

\begin{table}
\caption{Parameters of AGN jet}
\label{table:1}
\begin{tabular}{lll}
\hline\noalign{\smallskip}
Jet & life span [Myr] & Power [erg/s]  \\
\noalign{\smallskip}\hline\noalign{\smallskip}
1 & 25 & $10^{45}$ \\
2 & $2 \times 25$ & $5 \times 10^{44}$ \\
\noalign{\smallskip}\hline
\end{tabular}
\end{table}

To have a self-consistent model for jet power, one should assume
that an AGN's outbursts are conditioned by the accretion rate of
the black hole lurking at the center of the cluster. The
accretion rate will depend on the rate of inflow of matter toward the
cluster core due to radiative cooling loses, which form a self-sustained
loop \citep{BinneyTabor,Cattaneo}. Here we assumed fixed power and duration of the jet
based on observational estimates of known AGN jets (Table 1).

\section{Results}
\label{sec:2}

\begin{figure}[ht!]
\begin{center}
\includegraphics[height=2.5in, width=3.3in]{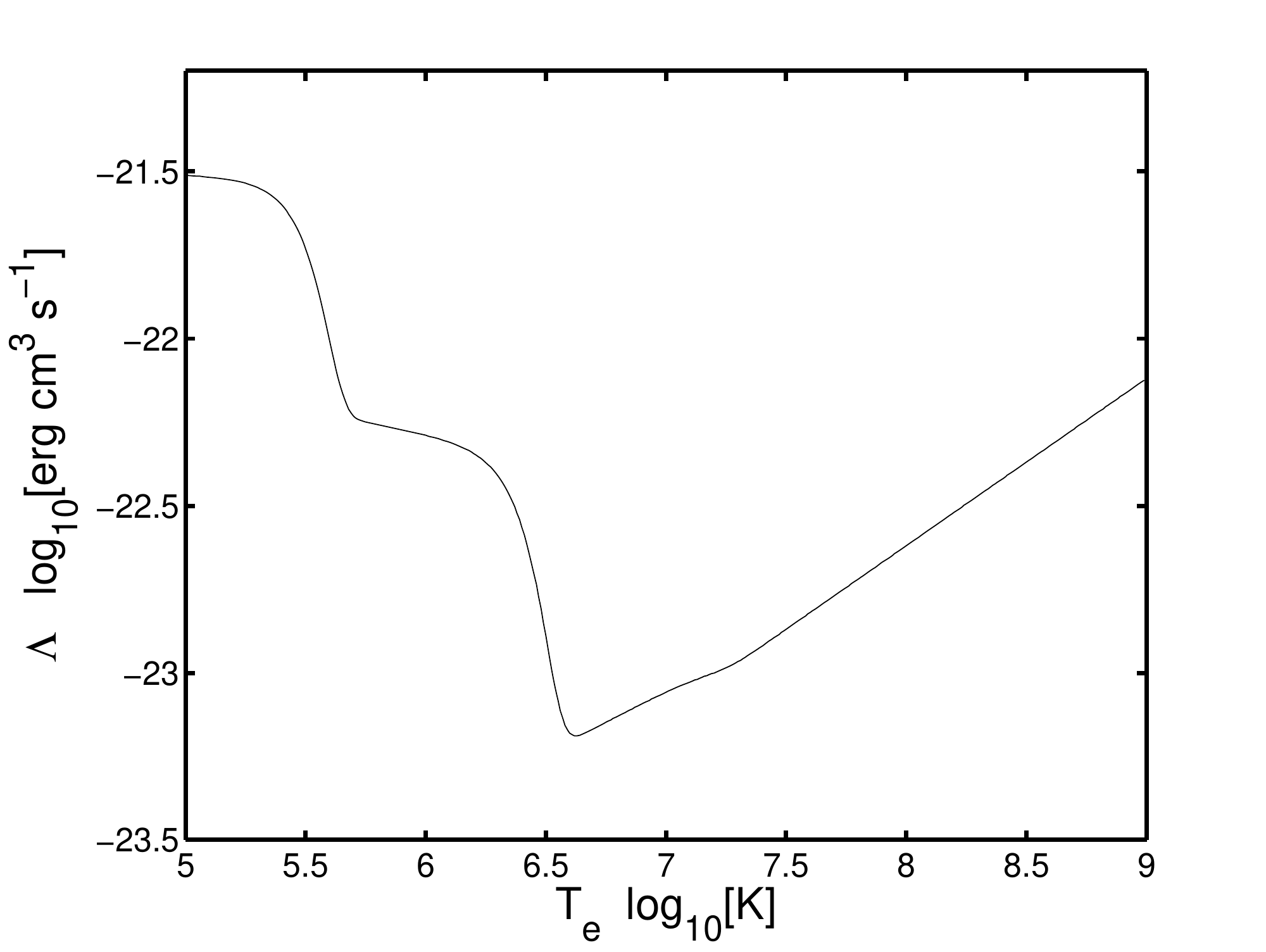}
\caption{Cooling function as function of gas temperature at 
$\frac{1}{2} z_{\odot}$.}
\label{fig:1}
\end{center}
\end{figure}

The cluster is allowed to cool passively for $3 \times 10^8\,$yr using line
and continuum radiation \citep{Sutherland}. The cooling function $\Lambda$ is
represented in (Fig.~\ref{fig:1}) as a function of the gas temperature for half solar
metalicity.

The rate of radiation of energy by a shell of radius r and width dr is:
\begin{equation}
j_{\rm rad}(r,t) = V_{\rm shell} (r,r + dr)n_e^2 (r,t){\Lambda (T_e (r,t))},
\end{equation}
 where $V_{\rm shell}$, $n_e$ and $\Lambda$ are the shell volume, electron
number density and cooling function (Fig.~\ref{fig:1}), respectively.  Under
the assumption that the plasma is optically thin, this energy is removed
from the computational domain. In the absence of AGN feedback, the system
suffers a cooling catastrophe after $3.7 \times 10^8\yr$.

\begin{figure*}
\includegraphics[height=3.5in, width=6.2in]{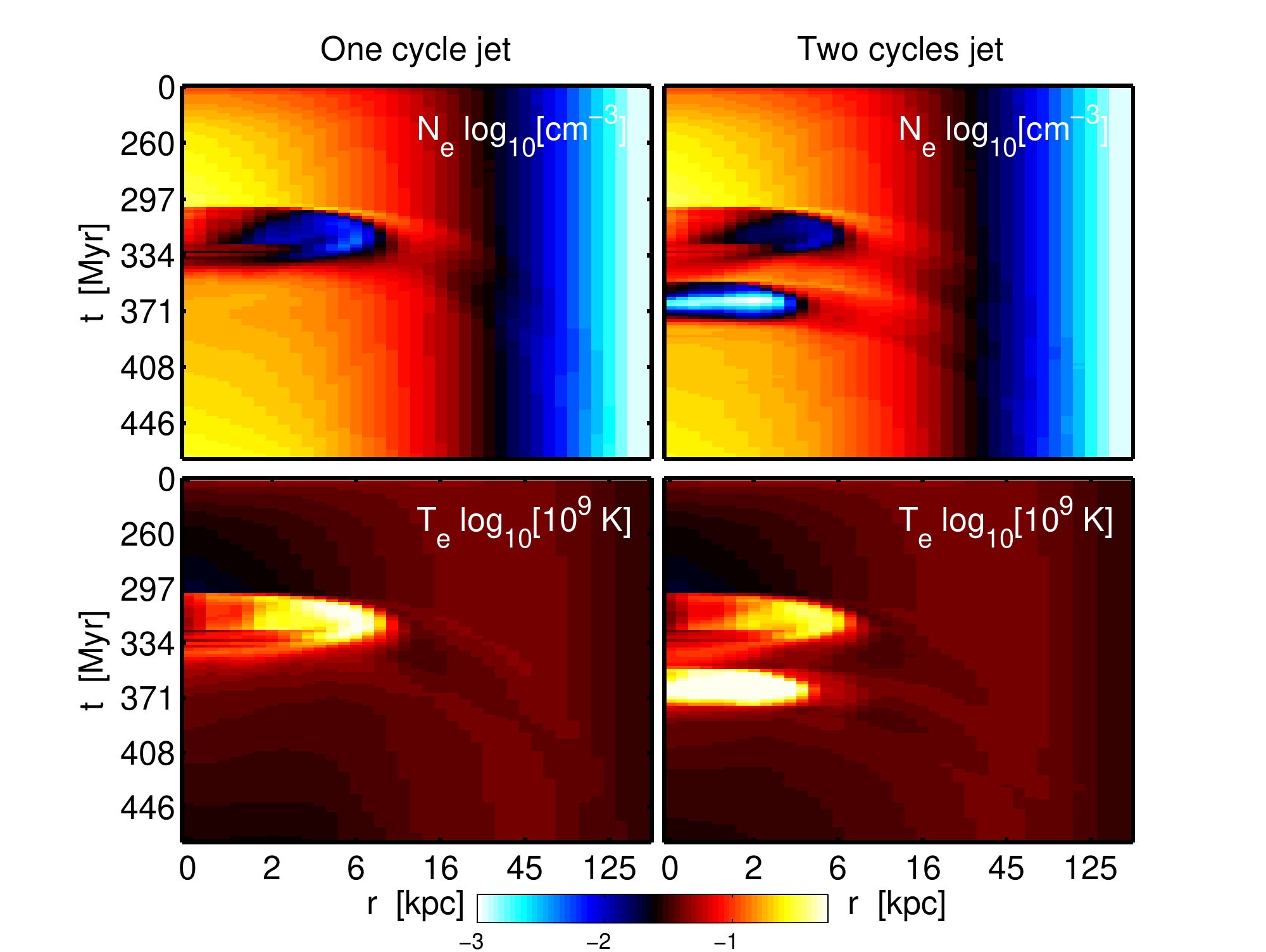}
\caption{Map of radial temperature and density profiles as function of time
for one cycle jet (left column) and two cycles jet (right column).}
\label{fig:2}
\includegraphics[height=4.5in, width=6.2in]{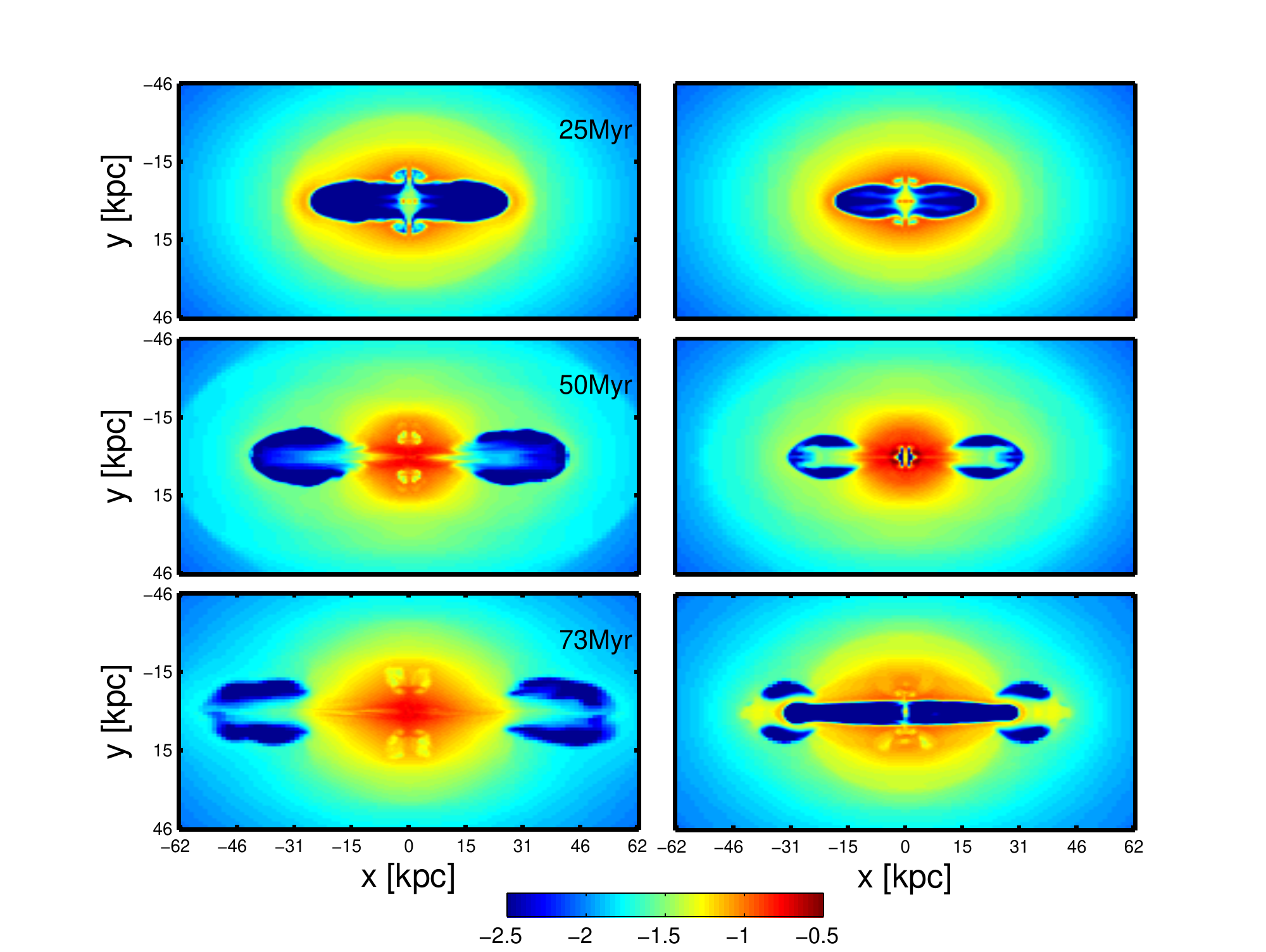}
\caption{Density map along the $xy$ plane ($x$ is the jet axis) and bubble
evolution, one cycle jet (left column); two cycles jet (right column). The
evolution of the bubble (dark blue region) at different times 25, 50 and
73\,Myr after jet ignition.}
\label{fig:3}       
\end{figure*}

Fig.~\ref{fig:2} shows $n_e$ and $T_e$ averaged over spherical shells of
radius $r$ as functions of $r$ and $t$. The left panel is for a simulation
in which there was only one outburst, while the right column is for a
simulation with two outbursts lasting 25\,Myr each, and separated by a quiescent
period of 25\,Myr (Table 1). 

\begin{figure}[ht!]
\begin{center}
\includegraphics[height=2.5in, width=3.3in]{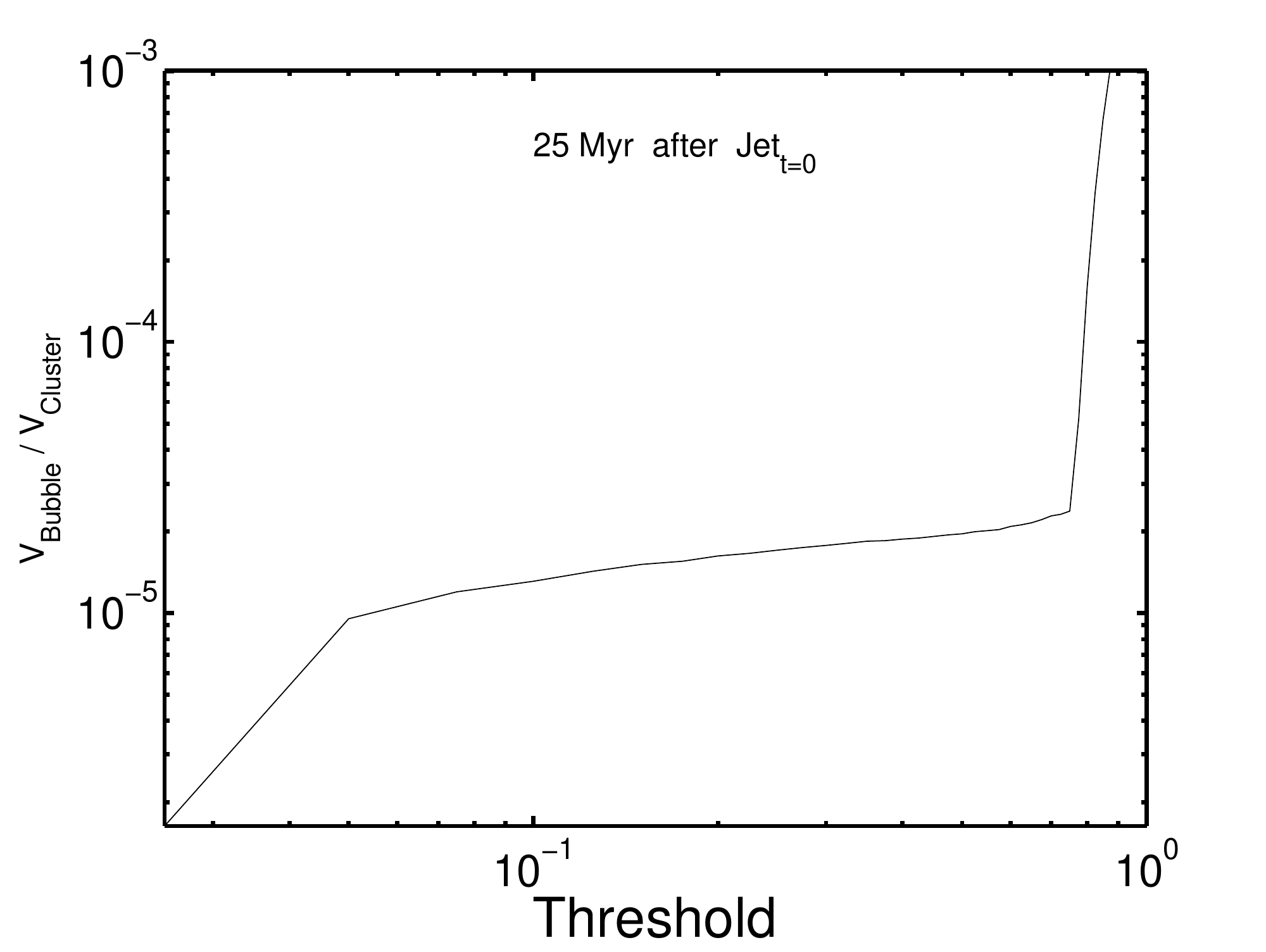}
\caption{The volume of the bubble as function of the density threshold.}
\label{fig:4}
\end{center}
\end{figure}

The upper row of Fig.~\ref{fig:2} shows that the density profile increases
to its maximum just before the jet ignites, because radiative losses cause
the central plasma to cool, contract and draw plasma in from outlying
regions. The jet immediately creates a low-density bubble (blue color near
the cluster core). This bubble then moves forward, driven by both jet
momentum and buoyancy (Fig~\ref{fig:3}). The bubble volume is determined by locating cells in which
the density is smaller than the averaged density at that
radius by a factor $f$.  Fig.~\ref{fig:4} shows the resulting volume as a
function of $f$, and on the basis of this plot we set $f=0.3$.

The temperature profiles in the lower
row (Fig.~2) show strong heating during the jet phase, and this heating can be
seen to offset the cooling process for about 50 Myr. Results not presented here show that short, powerful jets heat the gas
within 10 kpc of the cluster center less effectively than a longer-lived
lower-power jet, regardless of whether the latter has a single long spell of
activity ($\sim50\,$Myr) or has two 25-Myr blasts separated by a 25-Myr rest
period. The reason is that powerful jets (FR II type) tend to deposit their
energy far from the cluster core \citep{Omma}. Of the cases presented in
Fig.~2, the jet with two cycles couples more efficiently with the gas within
the inner cluster region. Although much of the gas uplifted during the first
cycle falls back during the $25\,$Myr jet quiescence, the bulk of the energy
injected in the second cycle passes to large radii along the channels that
the jet carved through the dense inner ICM during its first cycle (Fig.~3).

\section{Conclusion}
\label{sec:4}

From the measured bubble volumes and the pressure in the bubbles, we found
the energy associated with the bubbles, $ U_{\rm bubble} = pV+\gamma
pV/(\gamma-1)$, where $pV$ is the work done by the jet in creating the
bubble and $\gamma pV/(\gamma- 1)$ is the thermal energy within the bubble,
and $\gamma=5/3$ appropriate to a non-relativistic plasma. We found that
$U_{\rm bubble}$ is about 6 times less than the energy input from the jet
\citep{Binneyetal07}. Most of the energy not associated with bubbles is
divided between the kinetic and potential energy of the gas that is either
displaced by the bubbles or entrained in their wake. Some additional energy
has been thermalized in the shocks that move ahead of the bubbles.  The
kinetic and potential energy that surrounds the bubbles will eventually be
converted into heat through dissipation of the long-lived gravity waves in
which this energy is invested.

\section{Acknowledgements}
This work was supported by St Johns College, University of Oxford.

\label{lastpage}


\begin{thebibliography}{}

\bibitem[Basson \& Alexander(2003)]{BassonA}
Basson, J.F. \& Alexander, P., 2003, MNRAS, 339, 353

\bibitem[Binney(2005)]{BinneyRS}
Binney, J., 2005, Phil.\ Trans.\ R.\ S., 363, 739

\bibitem[Binney et al.(2007)]{Binneyetal07} 
Binney J., Alouani Bibi F., Omma H., 2007, MNRAS, 377, 142

\bibitem[Binney \& Tabor(1995)]{BinneyTabor}
Binney, J. \& Tabor, G., 1995, MNRAS, 276, 663

\bibitem[Birzan et al.(2004)]{Birzan}
Birzan L., Rafferty D.A., McNamara B.R., Wise M.W., Nulsen P.E.J., 2004.
ApJ, 607, 800

\bibitem[Blandford \& Znajek(1977)]{Blandford}
Blandford, R.D., Znajek, R.L., 1977, MNRAS, 179, 433

\bibitem[Br\"uggen \& Kaiser(2002)]{BruggenK}
Br\"uggen, M. \& Kaiser, C.R., 2002, Nat, 418, 301 

\bibitem[Bryan \& Norman(1997)]{Bryan}
Bryan G.L., Norman M.L., 1997, in ``Computational Astrophysics'', ASP Conf.\
Ser.\ 123, p.363

\bibitem[Carilli et al.(1994)]{Carilli} 
Carilli, C.L., Perley, R.A., Harris, D.E., 1994, MNRAS, 270, 173

\bibitem[Cattaneo \& Teyssier(2007)]{Cattaneo} 
Cattaneo, A., Teyssier, R., 2007, MNRAS, 376, 1547

\bibitem[Churazov et al.(2000)]{Churazov} 
Churazov E., Forman W., Jones C., B\"ohringer H., 2000, A\&A, 356, 788

\bibitem[Churazov et al.(2001)]{Churazov01} 
Churazov E., Br\"uggen M., Kaiser, C.R., B\"ohringer H., Forman W., 2001, ApJ, 554, 261

\bibitem[David et al.(2001)]{David} 
David L.P., Nulsen P.E.J., McNamara B.R., Forman W., Jones C., Ponman T. Robertson B., Wise M., 2001, ApJ, 557, 546

\bibitem[Fabian et al.(2001)]{Fabian} 
Fabian, A.C., Mushotzky, R.F., Nulsen, P.E.J., Peterson J.R., 2001, MNRAS, 321, L20

\bibitem[McNamara et al.(2000)]{McNamara} 
McNamara, B.R., Wise, M.W., Nulsen, P.E.J., David, L.P., Sarazin, C.L., 
Bautz, M., Vikhlinin, A., Forman, W.R., Jones, C., Harris, D.E., 2000, 
ApJL, 534, L135

\bibitem[Navarro et al.(1995)]{Navarro} 
Navarro, J.F., Frenk, C.S., White, S.D.M., 1995, MNRAS, 275, 720

\bibitem[Nipoti et al.(2005)]{NipotiBB}
Nipoti, C., Blundell, K.M \& Binney, J.J., 2005, MNRAS, 361, 633

\bibitem[Omma et al.(2004)]{Omma}
Omma, H., Binney, J., Bryan, G., Slyz, A., 2004, MNRAS, 348, 1105

\bibitem[Peterson et al.(2003)]{Peterson03}
Peterson et al., 2003, ApJ, 590, 207

\bibitem[Pizzolato \& Soker(2005)]{PizzolatoS}
Pizzolato, F. \& Soker, N., 2005, ApJ, 632, 821

\bibitem[Sutherland \& Dopita(1993)]{Sutherland}
Sutherland, R.S., Dopita, M.A., 1993, ApJS, 88, 253

\bibitem[Woodward \& Collela(1984)]{Woodward}
Woodward, P., Collela, P., 1984, J.Comp.Phys, 54, 115


\end{thebibliography}
\end{document}